\documentclass[paper,11pt,nopagenum]{AAS}		%

\usepackage{bm}
\usepackage{amsmath}
\usepackage[colorinlistoftodos]{todonotes}
\usepackage{subfigure}
\usepackage[colorlinks=false, pdfstartview=FitV, linkcolor=red, citecolor=red, urlcolor=red]{hyperref}
\usepackage{overcite}
\usepackage{xcolor}
\usepackage{tablefootnote}

\PaperNumber{26-027}

\begin{document}

\title{Distributed Observation Scheduling and Sensor Fusion for Large Agile Constellations}

\author{Nicholas Niziolek\thanks{Graduate Student, Autonomous Systems, University of Colorado Boulder}
    \ and
    Ryan Draves\thanks{Graduate Student, Autonomous Systems, University of Colorado Boulder}
    \ and
    Nolan Ales\thanks{Graduate Student, Autonomous Systems, University of Colorado Boulder}
    \ and
    Nolan Stevenson\thanks{Graduate Student, Autonomous Systems, University of Colorado Boulder}
    \ and
    Eli Jones\thanks{Student, Computer Science, University of Colorado Boulder}
    \ and
    Eric Frew\thanks{Professor, Autonomous Systems, University of Colorado Boulder}
    \ and
    Nisar Ahmed\thanks{Professor, Autonomous Systems, University of Colorado Boulder}
}

\maketitle{}

\begin{abstract}With the rapid increase in interest towards large scale earth observing satellite constellations, the problem of properly commanding and utilizing these systems has become increasingly relevant. This work focuses on
    a complete distributed approach to this problem, incorporating an observation task allocation and scheduling algorithm in addition to a sensor data fusion algorithm for end-to-end target coverage and processing. Observation task allocation and scheduling is performed in this work through the combination of the bid warped consensus based bundle algorithm (BWCBBA) and a lightweight observation scheduler for bid scoring. This method produces a unique mapping of tasks to command agents when allowed to iterate sufficiently. The approach to distributed data fusion in this work occurs primarily through an ownership model in which one satellite takes ownership of a system track in the constellation. After local information is shared between the sensing nodes to create the system tracks, the dedicated fusion nodes assume ownership for further information fusion and track information is disseminated throughout the constellation for data association. This paper analyzes the performance of these two algorithms over a set of increasingly decentralized architectures to highlight the performance losses compared to more traditional centralized methods.

\end{abstract}

\section{Introduction}

With the rapid proliferation of large scale earth observing satellite constellations, the need for autonomous scheduling, sensing, and data fusion has become increasingly relevant. These constellations can consist of hundreds of space assets, requiring extensive ground support to maintain continuous operation. This work focuses on a distributed autonomous end-to-end observation scheduling to target data association and fusion approach to track moving targets with limited prior information. The development of this distributed approach is primarily motivated by the need for robustness in operation, eliminating the single point failure mode seen with ground based control systems and allowing continuous operation of the constellation. Additionally, this distributed approach allows greater dynamism when handling space asset losses, as all information sharing and decision making is handled space-to-space without the need of ground detection and intervention.

The methodology implemented in this paper combines two main components: tasking and scheduling, and data association and fusion. The tasking and scheduling component of this research utilizes a decentralized, auction-based algorithm to distribute target ownership between teams of spacecraft. These teams consist of one command agent and a unique subset of all sensing spacecraft. Target allocation is done to simplify the scheduling problem onboard each spacecraft by limiting the set of considered targets, in addition to concretely defining ownership and responsibility for target observation. Auction bidding is scored using a lightweight scheduler in order to generate semi-realistic possible scores on each target in the current considered set. This approach allows for higher fidelity than simpler set-content-based heuristics, while retaining relatively low computational expense.

The data association and fusion component of this research uses distributed data fusion (DDF) techniques to decentralize target tracking across the constellation. Novel information is introduced into the system by spatially sharing local information, ensuring fusion satellites receive a complete enough picture to fuse measurements on the same target from multiple sources. Information associated to existing tracks is routed to the satellite that owns a known track.

Lastly, this work presents a simulation architecture that enables the research of these decentralized, space-based observation and tracking systems. We present the chosen system tradeoffs, fidelity capabilities, general design of this simulation, and the asynchronous event handling it emulates. We then used this simulation to produce experimental results of the observation schedule and sensor fusion algorithms in a large-scale constellation.

\section{Background}

\subsection{System Overview}

\begin{figure}[h!]
    \centering
    \includegraphics[width=0.45\textwidth] {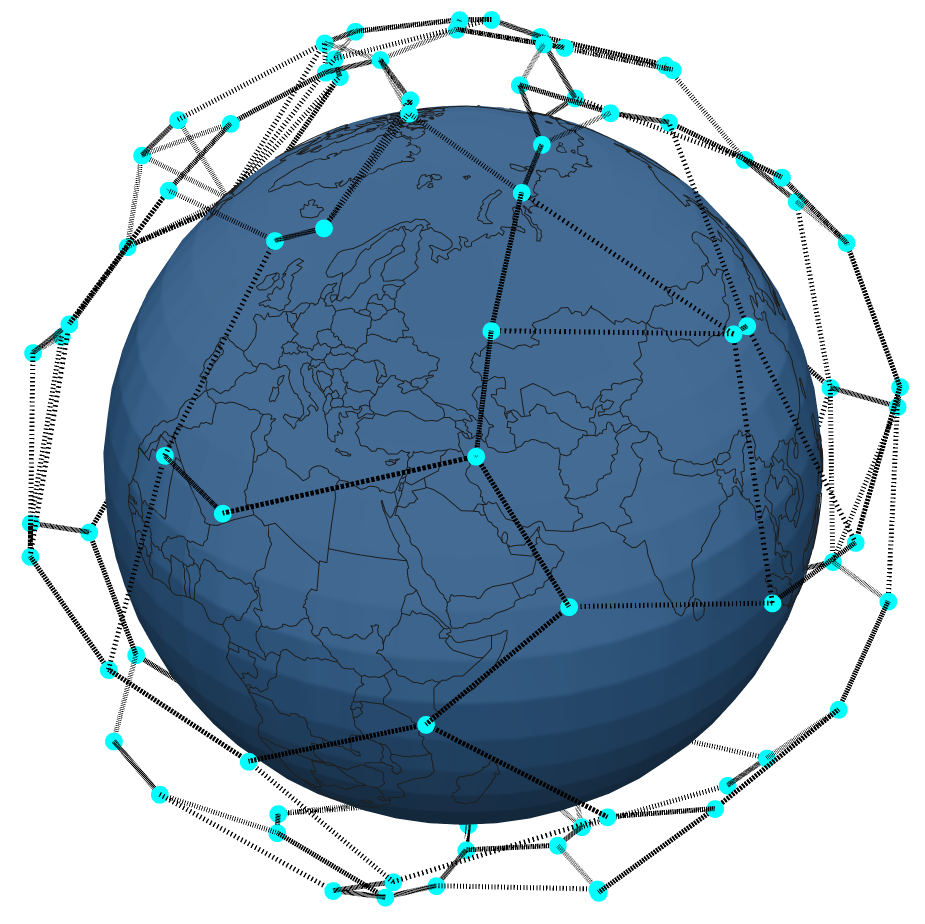}

    \caption{Reference observation constellation.}
    \label{fig:sat_comms}
\end{figure}

Our reference system for simulation experiments is a 108-satellite constellation in a low Earth orbit (LEO) Walker-Delta pattern. Satellites have inter-satellite communication links to exchange information in a mesh network as shown in Figure \ref{fig:sat_comms}. This work separates the notion of ``sensing satellites'', which are equipped with sensors and pointed to take angle measurements of observed targets, from ``fusion satellites'', which are responsible for the observation tasking of sensing satellites and data fusion of their measurements. For this work, sensing and fusion responsibilities are shared by each satellite in the constellation, which can be viewed as a single satellite or a pair of formation-flying satellites to the same effect.

Lastly, our system consists of a centralized ground asset that interacts with the constellation. When testing baseline scenarios of centralized estimation, this ground asset is responsible for collecting and processing all observations; when testing decentralized scenarios, the ground asset is a passive observer of the track information produced by the constellation.

\subsection{Decentralized Task Allocation}

The decentralized observation task allocation problem falls into the general multi-agent decentralized allocation problem set, which has many known solution approaches \cite{CBBA,GoodComp,CentDecentComp}. Multiple methodologies have been proposed to address the satellite constellation tasking problem with both market-based approaches \cite{CBBA_EOS,CentDecentComp} and optimization-based approaches \cite{DCOP}, the two main approaches seen in literature\cite{StateOfTheArt}. Market-based approaches make up the primary body of work in the satellite tasking field, being relatively simple in implementation and leveraging the large amount of work existing to perform auctions efficiently and with multiple types of scoring algorithms\cite{CBBA,BWCBBA}.

Much of the literature on decentralized satellite task allocation utilizes heuristic-based scoring functions, often based on problem geometry, to estimate observation quality\cite{CBBA_EOS}. These approaches produce quality solutions with low computational effort, but can suffer from a lack of fidelity in observation schedule feasibility. This drawback was the leading motivation in utilizing a lightweight scheduling algorithm able to provide higher fidelity approximations of realistic schedules, in addition to being significantly computationally lighter than a full fidelity scheduling algorithm.

\subsection{Observation Scheduling}

Observation scheduling involves generating a feasible motion plan for each sensing spacecraft considered in order to best cover the given set of tasks. The approach used in this paper builds and solves a mixed integer linear program (MILP) to generate a set of discrete pointing directions for a given discrete time vector. Satellite scheduling through the use of MILPs is seen throughout the observation scheduling field, many showing very high fidelity and tied heavily to EOS dynamics \cite{EOS_Optim,StereoAndSlew}, with continuous pointing and slew times. This paper uses principles from these models through a lower fidelity and significantly higher discretized approach, relying on selection from a set of pointing directions computed during preprocessing \cite{discpoint}. This approach maintains some of the fidelity gained by modeling a sensor with finite viewing capabilities and the relative performance of sensing opportunities, while not considering the full dynamics, and therefore making the problem significantly harder to solve.

\subsection{Distributed Data Fusion}

The estimation of moving targets from satellite sensor sources is a multi-target tracking (MTT) and data fusion problem. Bar-Shalom et al\cite{bar2011tracking} provide a detailed review of common algorithms to address the core challenges of MTT. \textit{Track-before-detect}, for instance, relates to the initialization of tracks before declaring a detected target and is approached with solutions such as particle filters or multi-hypothesis trees\cite{daveyComparisonDetectionPerformance2007}. Relating measurements to existing tracks or potentially new tracks is an issue of \textit{data association} and is approached with methods such as hypothesis trees, probabilistic data association filters\cite{1145560}, or simple global nearest neighbor (GNN) assignments. The choice of the filter itself depends on the tracking needs, with common methods such as Kalman filters, particle filters, or Gaussian mixture models (GMMs). \textit{Track-to-track fusion} deals with identifying and resolving redundant tracks, where a redundant track is when multiple tracks have been created over the same target; this can occur when the track initialization mistakenly splits independent measurement sources over the same target as unique tracks. Covariance intersection is a common choice for track-to-track fusion, as is the straightforward but computationally intensive option of rewinding a track through its entire measurement history.

Diffusing the multi-target tracking problem across a network of computing nodes, such as a satellite constellation, is an application of distributed data fusion (DDF). This creates additional challenges for information sharing across nodes to create a consistent set of system tracks. A notable problem is the issue of ``rumor propagation'', where nodes share common information and thus duplicate the information they fuse\cite{campbellDistributedDataFusion2016}. In the case of a constellation, where inter-satellite communication links form an ad-hoc, dynamic network topology, it becomes trivial to duplicate information in a DDF approach. Approaches to this problem involve either the prevention of common information sharing, such as passing messages along a singly connected graph, or reconciling duplicate information at a later time. In addition to common information sharing, processing local information in DDF can create additional redundant tracks over the same targets. For instance, if two sensors observe the same target but have not shared this information with a common node, two tracks with unique information may be formed over the same target.

\subsection{Constellation-Based Target Tracking}

Previous work on constellation-based target tracking has explored various tracking architectures primarily in space-to-space applications. Several works focusing on decentralized estimation techniques have explored consensus-based Kalman filtering methods in applications with small-scale constellations or formation flying\cite{doi:10.2514/6.2015-1553,jiaCooperativeSpaceObject2016,GONG2023108660}. Federated architectures have also been explored in formation flying for the purpose of tracking nearby space objects\cite{haysStateOmniscienceCooperative2024,haysAutonomousLocalCatalog2023}.

Gao et al.\cite{gaoOnboardInformationFusion2023} provide a detailed review of space-based Earth observation applications and practical considerations for conducting information fusion onboard satellites. Whereas information fusion is traditionally performed on the ground by means of downlinking information to a centralized ground asset, onboard fusion poses advantages for the resiliency and responsiveness of Earth observation applications, such as time-critical surveillance systems. New challenges are posed, however, by the general SWAP constraints of onboard compute and thus drive the need for novel information fusion techniques for these domains.

\section{Methodology}

\subsection{Simulation Architecture}

In addition to algorithmic contributions, we present a simulation architecture created to enable medium-fidelity investigations of a decentralized space-based system. When modeling a large decentralized system, the fidelity of the system is impacted by the underlying architecture. A synchronous simulation, for instance, will typically not properly account for the processing and communication delays inherent in any real-world system, nor the desynchronization inevitable to a global-scale sensor network. Likewise, a naive design allocating a subprocess or thread to each satellite introduces starvation and other timing inconsistencies that are difficult to avoid without expensive computational resources.

\begin{figure}[h!]
    \centering
    \includegraphics[width=1.0\textwidth] {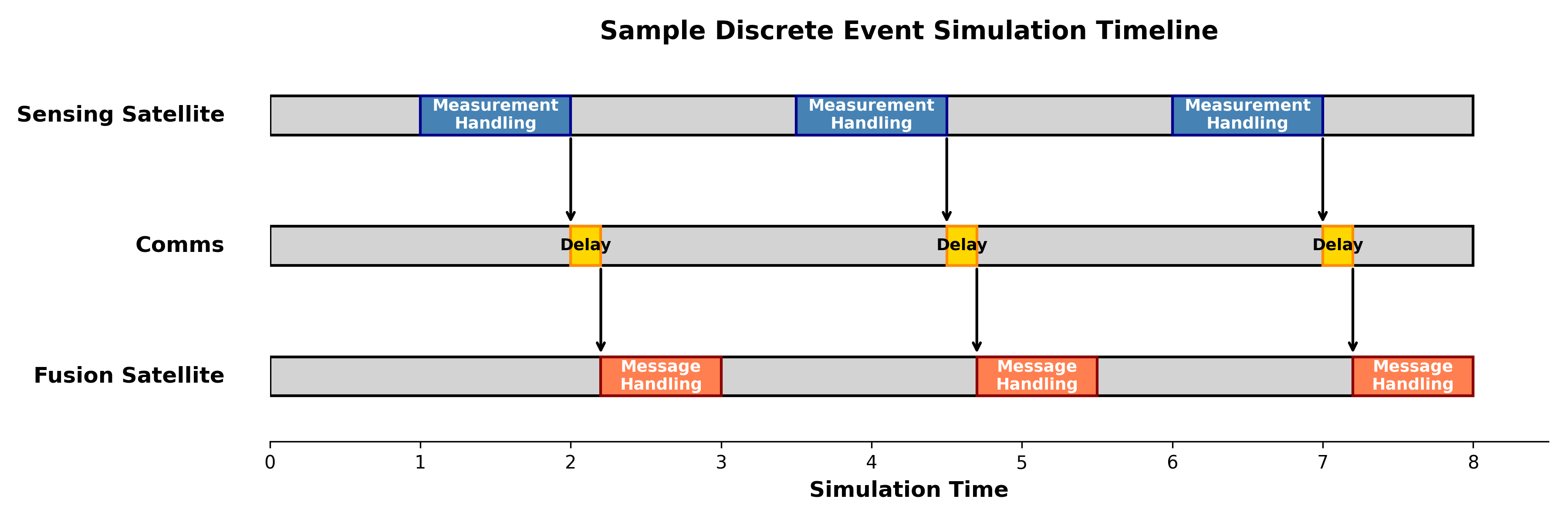}
    \caption{Sample discrete event flow between satellites.}
    \label{fig:des_timeline}
\end{figure}

To this end, we model the environment using Discrete Event Simulation (DES), where individual satellites are given their own \textit{process} to wait for and handle events. A \textit{process} moves its own clock forward as it handles and waits for events, giving the satellites their own simulated clock time to reference. A centralized coordinator manages the DES processes to ensure events are handled across the simulation in chronological order. Figure \ref{fig:des_timeline} shows a simplified example of a sensing satellite gathering periodic measurements and a fusion satellite processing these measurements after accounting for the speed of light and transmission delay in the inter-satellite communications. This architecture allows for the processing and communication delays to be tuned arbitrarily, enabling asynchronous interactions across the constellation that closely follow the timing patterns of a real-world system while preserving reproducibility in the simulation results.

\begin{figure}[h!]
    \centering
    \includegraphics[width=0.75\textwidth] {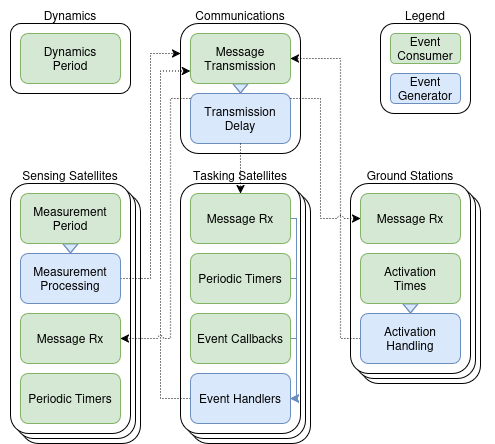}

    \caption{Event flow architecture of the Discrete Event Simulation.}
    \label{fig:des_architecture}
\end{figure}

\paragraph{Processes}
The simulation is modeled such that each satellite and ground station has its own DES process, and a communications process acts as an intermediary to account for delays in inter-satellite messages. A separate dynamics process updates the communication links between the satellites at a coarser interval than position updates. Figure \ref{fig:des_architecture} shows the event interactions of these processes and the types of events each process waits upon.

\paragraph{Timing}
To account for processing delay, each event handling must be given an amount of simulated delay to represent the duration of the simulated process. To create a reproducible simulation with reasonable timing fidelity, we profile each event handler and apply a linear fit to its real-time duration. For example, processing measurements may be found to take approximately 2 milliseconds per measurement plus a flat overhead cost of 3 milliseconds.

\subsection{Decentralized Observation Task Allocation}
\label{sec:tasking}

Observation task allocation and scheduling is accomplished in this work through the combination of a lightweight observation scheduling algorithm and an auction based task consensus algorithm \cite{TaskingIEEE}. The architecture of this system is presented as the interaction between commanding agents (fusion satellites) and sensing satellites. Each command agent is assigned a unique subset of the total set of sensing satellites, referred to as a team. Command agents are responsible for scheduling the observations of their team members and communicating with other command agents to deconflict task assignments. Deconfliction between command agents ensures that tasks are owned only by the commander able to bid highest.

In this work, teams are formed by decomposing the set of satellites by orbital plane, and dividing each plane into two parts using an alternating selection scheme\cite{DCOP}. This method maximizes continuous coverage of targets for each team, and provides relatively high overlap between teams occupying the same orbital plane. This aids inavoiding over-tasking one specific team if the problem geometry results in that team providing most or all of the visibility on a set of targets.

In Figure \ref{fig:DAS_Flow} the full system architecture is shown. In the top bar, simulation parameters are passed to the fusion satellites, and simulated potential measurements are passed to the sensing satellites. These potential measurements are a set of on-target and false positive measurements, containing information on time of observation, in-track bearing, cross-track bearing, and satellite making the observation. These measurements are generated assuming a field of view (FOV) covering all visible area of Earth below. These are filtered by the sensing satellites in the left most box on the figure after receiving schedules from the command agents (fusion satellites). The fusion satellites are shown to perform several steps prior to broadcasting schedules, primarily preprocessing and consensus. Preprocessing is run once per planning horizon, and the resulting data is fed into the scheduler model to inform observation scheduling.

\subsection{Tasking Preprocessing}

\begin{figure}
    \centering
    \includegraphics[width=1.0\linewidth]{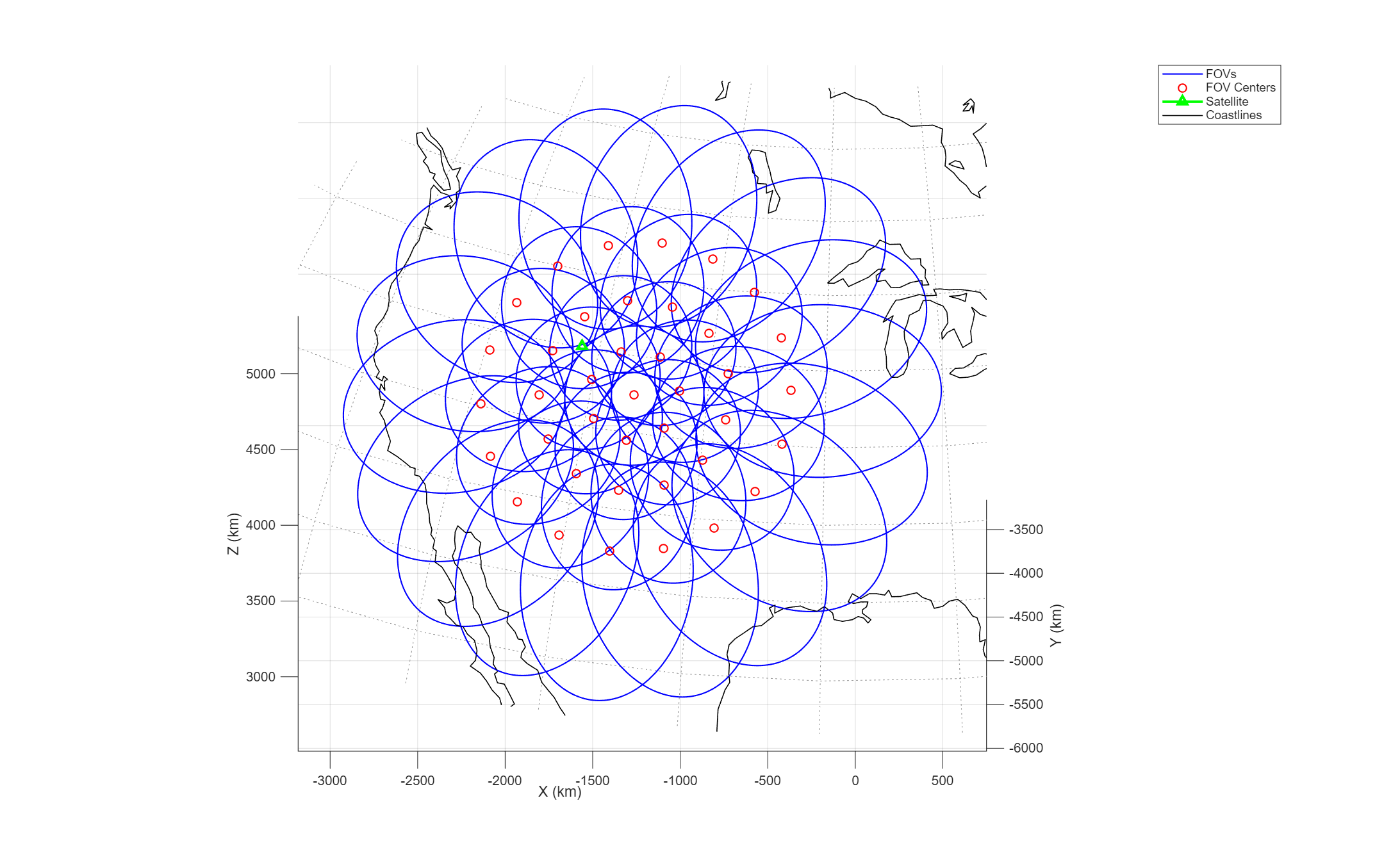}
    \caption{Discretized FOV options.}
    \label{fig:FOV}
\end{figure}

\begin{figure}[h!]
    \centering
    \includegraphics[width=1.0\linewidth]{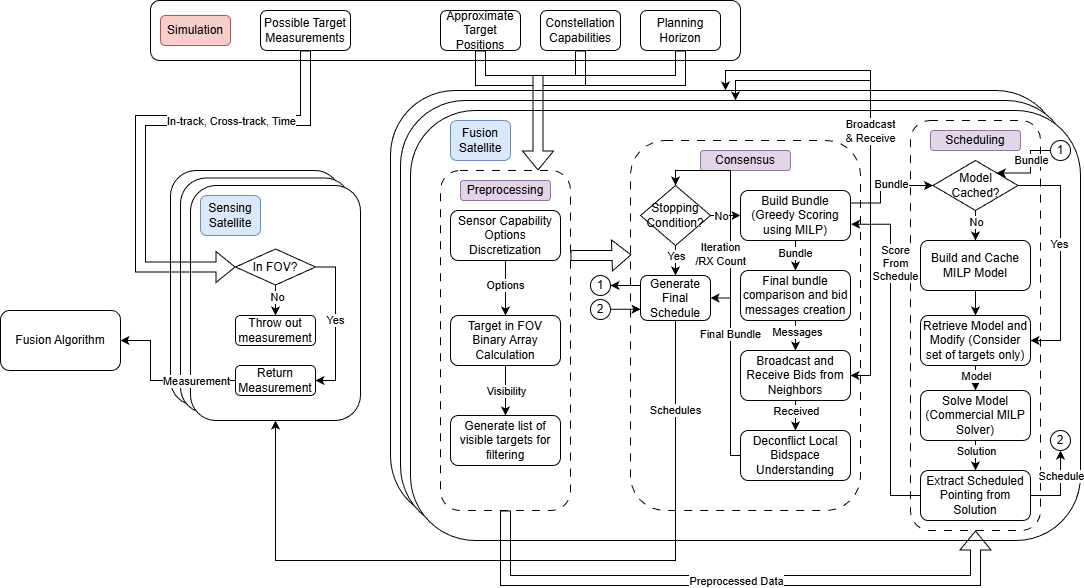}
    \caption{Decentralized allocation and scheduling architecture.}
    \label{fig:DAS_Flow}
\end{figure}

Preprocessing is required to alter the highly non-linear observation scheduling problem into a discretized linear model. In the first preprocessing block seen in Figure \ref{fig:DAS_Flow}, the sensing satellite capabilities are discretized into a set of options at each discrete time step. These options can be any set of pointing directions, but for this work are assumed to be a set which fully covers the visible ground below the satellite. The discretization\cite{discpoint} can be seen in Figure \ref{fig:FOV}, where the blue ellipses show a circular FOV projected onto Earth's surface from a satellite 1500km in altitude with a thirty degree FOV as is used below. As the FOV becomes more off-nadir, the ellipses elongate, and cover larger areas.

The second preprocessing block on the diagram takes these options as input to generate a binary array of satellite-target-time-option visibility checks, where each 1 entry represents a target visible in a satellite's FOV at a specific time looking in a specific direction. This check is done through comparing the orientation of each FOV surface normal with the vector from satellite to target. If the target is visible, all alignments should be positive. This is shown in Equation \ref{eq:vis} where $\vec{v}_i$ represents the ith FOV edge vector and $\mathbf{R}_S^I$ represents the rotation matrix to transpose the resulting vector from the sensor frame (S) to the inertial frame (I) based on the current orientation. The calculated visibilities are stored in a binary array (g) indexed by the satellite (s), time (k), target (j), and option (n).

\begin{equation}
    \vec{r}_{ts}\cdot \mathbf{R}_S^I(\vec{v}_i\times \vec{v}_{i+1}) \ge 0 \; \forall i
    \label{eq:vis}
\end{equation}

The final block represents a filtering step to limit the targets each commander considers to only the set of those visible at some point in time to at least one team member.

\subsection{Lightweight Observation Task Scheduling}

A lightweight observation scheduling algorithm is used to score bundles of tasks by creating realistic potential observation schedules for each sensing satellite in the command agent's team. The scheduler considered in this work is formulated as a mixed integer linear program (MILP). The goal of this optimization model is to maximize the number of observations per target, with a diminishing returns based scoring method.

\begin{equation}
    \mathbf{X}^* = \arg \underset{\mathbf{X}}{\text{max}}\quad \sum_{T} r_j(\tau_j(\mathbf{X}))
    \label{eq:obj}
\end{equation}

Where $\mathbf{X}^*$ represent a binary array indexed by spacecraft (s), time (k), and pointing option (n). This objective is subject to the physical constraint that a spacecraft must point in only one direction at any given time step, enforced by Equation \ref{eq:constr}.

\begin{equation}
    \sum_n x_{s,k,n} =1 \quad \forall \quad s\in S, k \in K.
    \label{eq:constr}
\end{equation}

The quality of the schedule is driven by the observation count $\tau_j$, which represents the number of times target $j$ is successfully observed across the constellation where $g$ represents the binary visibility matrix, and $x$ represents a single entry from the schedule $\mathbf{X}$ (Equation \ref{eq:count}).

\begin{equation}
    \tau_j = \sum_{S} \sum_{K} \sum_{n \in \theta_{s,n}} g_{s,k,j,n}*x_{s,k,n} \quad \forall \quad j \in T
    \label{eq:count}
\end{equation}

The task reward is a piecewise linear function, designed to simplify computation in the MILP solver and capture the diminishing returns nature of repeated observations (Equation \ref{eq:reward}). The constant arrays $Y$ and $X$ represent the values and break points respectively.

\begin{equation}
    \begin{split}
         & r_j (\tau) =                                                                                                                   \\
         & w_j \cdot \begin{cases}
                         Y_1+\frac{Y_2-Y_1}{X_2-X_1}(\tau-X_1)         & \tau \le X_1             \\
                         Y_i+\frac{Y_{i+1}-Y_i}{X_{i+1}-X_i}(\tau-X_i) & X_i \le \tau \le X_{i+1} \\
                         Y_m+\frac{Y_m-Y_{m-1}}{X_m-X_{m-1}}(\tau-X_m) & \tau\ge X_{m}
                     \end{cases} \\&\quad \forall\; j \in T.\\
    \end{split}
    \label{eq:reward}
\end{equation}

For simplicity, the MILP solver is represented as a callable function, where the input is the set of tasks, satellites, time steps, and options, and the output is the resulting schedule ($\mathbf{X}$) and the score ($R$) (Equation \ref{eq:scheduler}).

\begin{equation}
    \{\mathbf{X}, R\} = \text{MILP}(T,S,K,\Theta)
    \label{eq:scheduler}
\end{equation}

\subsection{Bid Warped Consensus Based Bundle Algorithm}

The task consensus and allocation framework used in this work is the bid warped consensus based bundle algorithm (BWCBBA)\cite{BWCBBA}. This algorithm is a modification of the consensus based bundle algorithm (CBBA)\cite{CBBA} which was made to assign bundles of tasks to a set of agents. This algorithm works through three main phases: bundle formation, bid communication, and deconfliction. During the first phase, bundle building, each agent greedily adds tasks to it's bundle, adding only those able to beat bids placed by other agents. Tasks are scored using the observation task scheduling algorithm described above, where the score is based on mission coverage of the tasks contained in the bundle. The score for individual tasks is extracted from the total score as the marginal change in score (Equation \ref{eq:score}). Due to the nature of MILP solutions, this score is not strictly submodular, and thus is warped for BWCBBA using Equation \ref{eq:BW} to maintain consensus properties.

\begin{equation}
    R_{i,j} = \text{MILP}_{i}(b_i \oplus j)-\text{MILP}_{i}(b_i).\\
    \label{eq:score}
\end{equation}

\begin{equation}
    \bar{R}_{i,j} = min\{R_{i,j},\bar{R}_{i,j_{|b_i|}}\}
    \label{eq:BW}
\end{equation}

During the second phase, bid communication, each agent generates bids on each task change in their bundle, formulated as either update bids or drop bids (Equation \ref{eq:bids}). Bids contain the identity of the agent placing them (i), the identity of the task being bid on (j), and the warped bid score ($\bar{R}_{ij}$) (Equation \ref{eq:bid}). These bids are broadcast to other command agents over the available communication network. Received bids are added to each agent's understanding of the global bid space, and conflicts are resolved by winning bids or tie breaking. This algorithm iterates until a stopping condition is met.

\begin{equation}
    \bar{s}_{ij} = \langle i,j,\bar{R}_{ij}\rangle
    \label{eq:bid}
\end{equation}

\begin{equation}
    \begin{split}
        \text{drop bid}   & = \langle \bar{s}_{ij},-1,k \rangle \\
        \text{update bid} & = \langle \bar{s}_{ij},1,k\rangle
    \end{split}
    \label{eq:bids}
\end{equation}

For this work, two stopping conditions are used. The original work on the CBBA required each agent to iterate until no new messages were received for a set number of iterations. In this work, this condition requires too much computation time, and thus commanders would not be able to terminate consensus in time to perform scheduling for the planning horizon. Instead, each agent iterates until one iteration without received messages, or until an iteration limit is met. In practice, this still generally results in few to no conflicts in the final assignment.

\subsection{Distributed Data Fusion}

For the fusion architectures, assume that the tasking and pointing of satellites has already been completed, making the resultant measurements the beginning of the system's information flow. Each architecture presented will also use the following linear dynamics model and nonlinear measurement model with additive white Gaussian noise to capture the target state. The nonlinear measurements are the in-track and cross-track angular measurements of observed targets from the sensor's perspective.

\begin{align}
    x_{k+1} & = F x_{k} + w_{k}, \quad w_{k} \sim \mathcal{N}(0, Q) \\
    z_k     & = h(x_k) + v_k, \quad v_k \sim \mathcal{N}(0, R)
\end{align}

\begin{figure}[h!]
    \centering
    \includegraphics[width=0.75\textwidth] {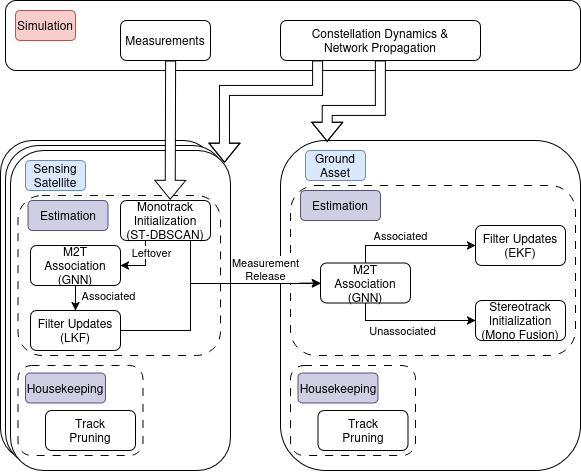}

    \caption{Centralized estimation architecture.}
    \label{fig:centralized_estimation}
\end{figure}

\paragraph{Centralized Estimation}
We first outline a centralized architecture described in Figure \ref{fig:centralized_estimation}. As it operates on the global set of measurement information, this architecture serves as a practical upper bound to compare the track quality of our decentralized method against. In this model, sensing satellites are responsible for ferrying measurements down to a centralized ground asset. The central ground asset completes the multi-target tracking system by performing data association and filter updates with the ingested measurement information. Fusion satellites bear no responsibilities apart from transporting multi-hop messages from sensing nodes and are thus omitted from the diagram.

Rather than sending bandwidth-intensive raw measurement data to the central ground estimator, the sensing satellites first solve the track-before-detect problem by initializing \textit{monotracks}, or single-observer tracks, from the raw measurements. ST-DBSCAN \cite{BIRANT2007208} is used for spatial-temporal clustering of measurements across a time horizon for initialization. Leftover measurements are then associated with global nearest neighbor (GNN) association to the existing set of monotracks, then updated using a linear Kalman filter about the in-track and cross-track bearings of the sensing satellite. Associated measurements are then ferried to the central ground asset.

From the ground, a similar routine is performed to estimate tracks from multiple views, or \textit{stereotracks}. GNN association of measurements to stereotracks occurs first in the pipeline, with extended Kalman filter updates on the associated measurements. Unassociated measurements are passed to a stereotrack initialization algorithm in which monotracks from multiple satellite sources are associated together with another GNN association and estimated with nonlinear least squares.

\paragraph{Decentralized Estimation}
When distributing estimation across the constellation, an architecture is driven by the approach to coalescing multiple sources of information (different satellites' measurements) to a single satellite. We outline a decentralized approach where the initial fusion of information occurs by having sensing satellites geocast their measurements. This solves the coalescing problem by ensuring at least one fusion satellite has received multiple sources of information on the same target, but in doing so, introduces duplicate information into the system that must later be reconciled.

\begin{figure}[h!]
    \centering
    \includegraphics[width=1.0\textwidth] {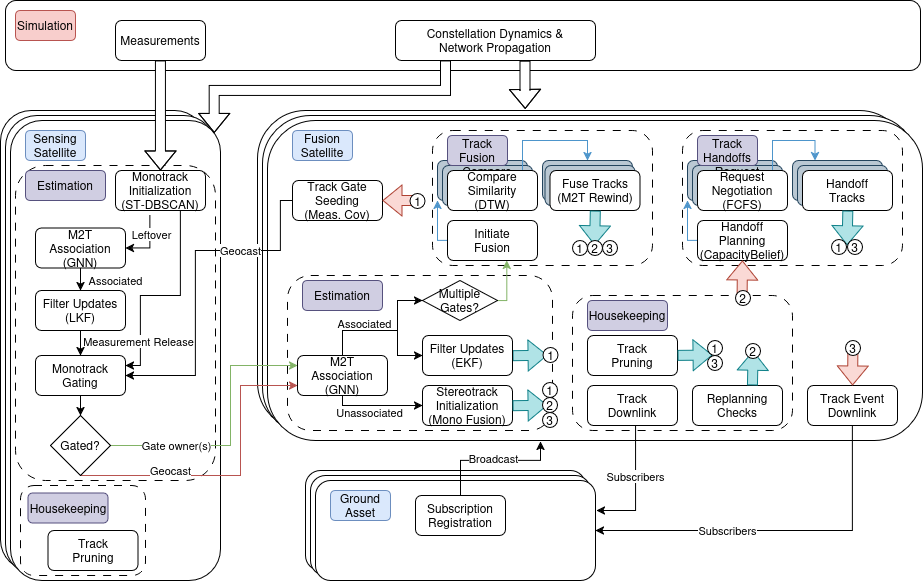}

    \caption{Decentralized estimation architecture.}
    \label{fig:decentralized_estimation}
\end{figure}

In Figure \ref{fig:decentralized_estimation}, we first alter the scheme from the centralized approach by passing released measurements through gates before being shared from the sensing layer to the fusion layer. Existing stereotracks from the fusion layer are disseminated as gates in the frame of nearby sensors. Released measurements within one or more gates are shared with the owning fusion satellite, while ungated measurements are geocast into the fusion layer.

In the fusion satellite layer, the same estimation algorithms run on the centralized ground asset are now run on each fusion satellite using the measurement information it has received. The track fusion module handles the reconciliation of duplicates by detecting tracks with overlapping measurement gates, comparing them with Dynamic Time Warping, particularly FastDTW\cite{doi:10.3233/IDA-2007-11508}, and fusing duplicate tracks with measurement rewinding. The track handoff module is responsible for planning track ownership transfers across the constellation to balance computational and bandwidth loads while keeping track ownerships close to the target being tracked. This planning is performed in a decentralized manner by using approximate inference over the satellite constellation state to inform a handoff policy; this work is explored in depth in Draves et al\cite{doi:10.2514/6.2026-1993}.

Lastly, in the ground layer, we use a simplified model for downlinking data out of the fusion layer and into a layer of ground assets. Ground assets are modeled as subscribers to regions of interest and register their subscription to the fusion layer.

\section{Results}
To analyze the system, we compare four combinations of tasking and fusion architectures. Our first test case omits tasking to examine baseline fusion performance. Satellites instead point nadir and are given a FOV to Earth's horizon, which is approximately a $54^\circ$ half angle from their orbit altitude. The remaining test cases mix and match the centralized and decentralized architectures for the tasking and fusion systems, with satellites given a $30^\circ$ full angle square FOV and the same sensor noise model as the first test case. The smaller FOV provides less than $10\%$ of the Earth surface coverage as the full horizon FOV; this limits the theoretical best result when comparing to the fusion-only scenario.

For each test case, we simulate a 30-minute scenario in which a few dozen targets are introduced into the environment in waves.

\subsection{Tasking Performance}

Tasking performance is evaluated by comparing centralized and decentralized architectures. The centralized configuration applies the lightweight scheduler to all sensing spacecraft and targets, representing an upper bound on performance with full coordination authority. For decentralized configurations, we measure mission performance per auction iteration, the number of task ownership conflicts, bid performance, and the number of messages propagated through the network.

In the selected scenario, a maximum of 64 targets are instantaneously observable to any of the 108 sensing spacecraft. This spacecraft-to-target ratio of 1.7:1 results in excessive sensor density for the given sensing problem. Additionally, this scenario exhibits greatly clustered targets (maximum coefficient of variation for five degree latitude and longitude grid cells of 1.4), causing a large inequality in required sensing activity spatially over the constellation. This results in a poor setup for single target ownership, as some observation opportunities will be lost from neighboring teams.

\begin{figure}
    \centering
    \includegraphics[width=1.0\linewidth]{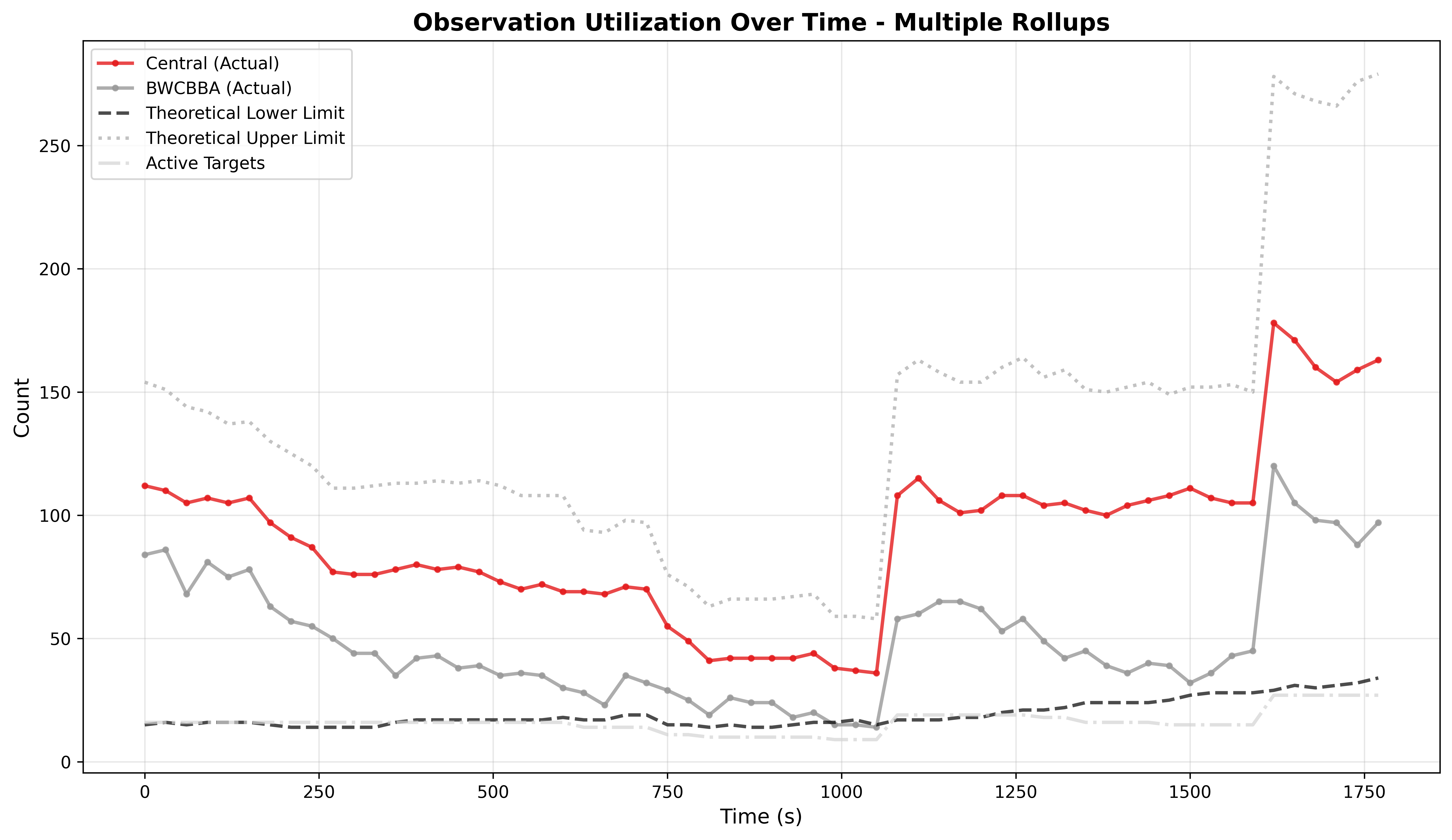}
    \caption{Observation utilization vs. time comparison.}
    \label{fig:obsv}
\end{figure}

\begin{figure}
    \centering
    \includegraphics[width=1.0\linewidth]{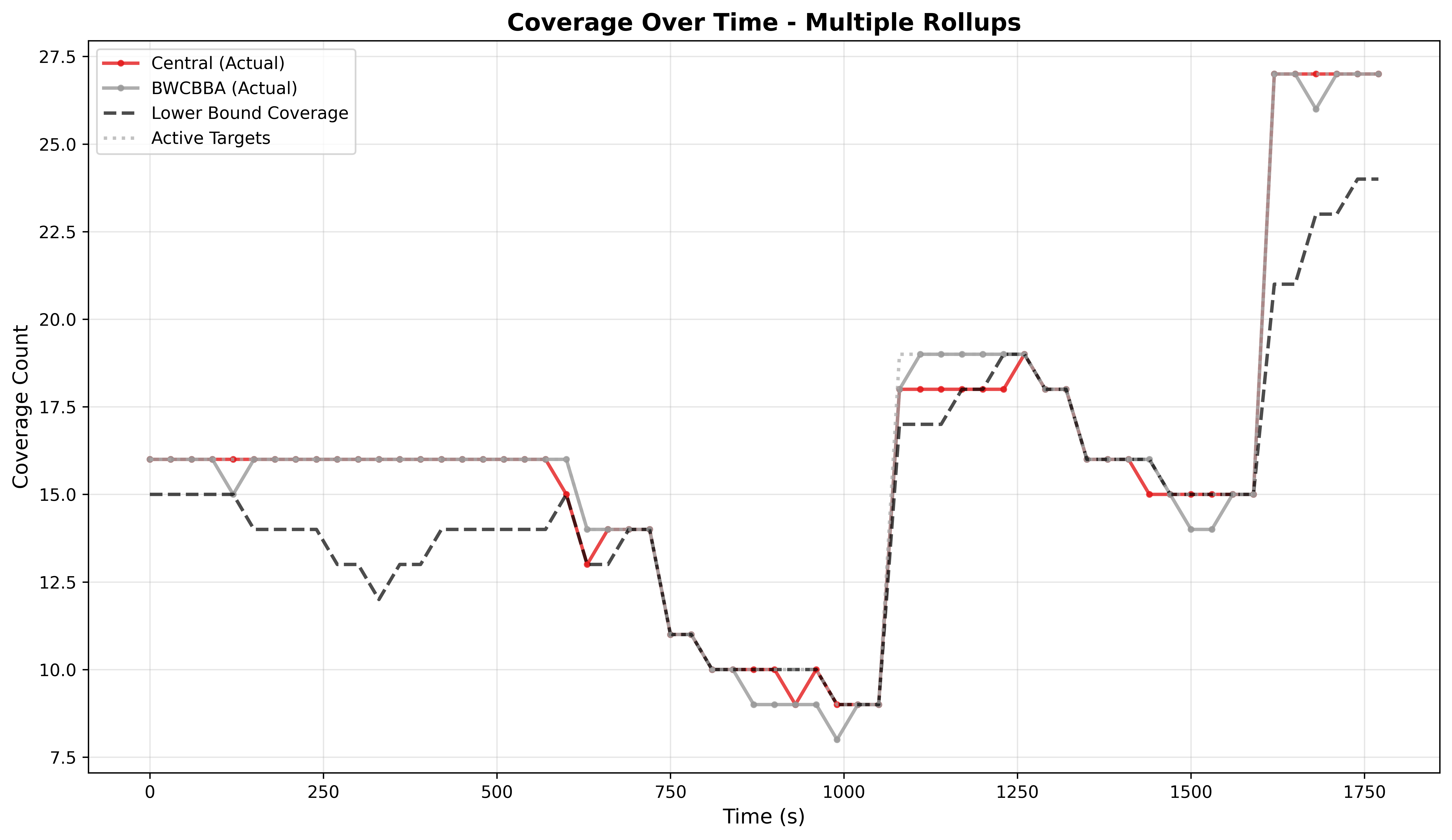}
    \caption{Coverage vs. time comparison.}
    \label{fig:cover}
\end{figure}

Figures \ref{fig:obsv} and \ref{fig:cover} compare decentralized and centralized performance. Figure \ref{fig:obsv} shows observation utilization between theoretical bounds: an upper bound assuming all visible targets are observable simultaneously, and a lower bound of one target per satellite. The centralized approach achieves roughly 50\% of maximum utilization while decentralized reaches roughly 25\%. Both exhibit declining trends due to target dispersion over time where targets initially clustered in tight groups spread apart, reducing simultaneous observability. The gap between both algorithms can be primarily explained by lost observation opportunities due to single target ownership.

Figure \ref{fig:cover} shows target coverage (number of targets with at least one observer). Despite the utilization gap, both approaches provide similar coverage, with most targets observed at each tasking timestep. It can be noted that the coverage occasionally drops below that of the lower bound. This can be explained by the incomplete edge-of-horizon coverage the sensing discretization provides (Figure \ref{fig:FOV}), whereas the lower bound considers all targets above or on the horizon.

\begin{figure}
    \centering
    \begin{minipage}{.5\textwidth}
        \centering
        \includegraphics[width=1\linewidth]{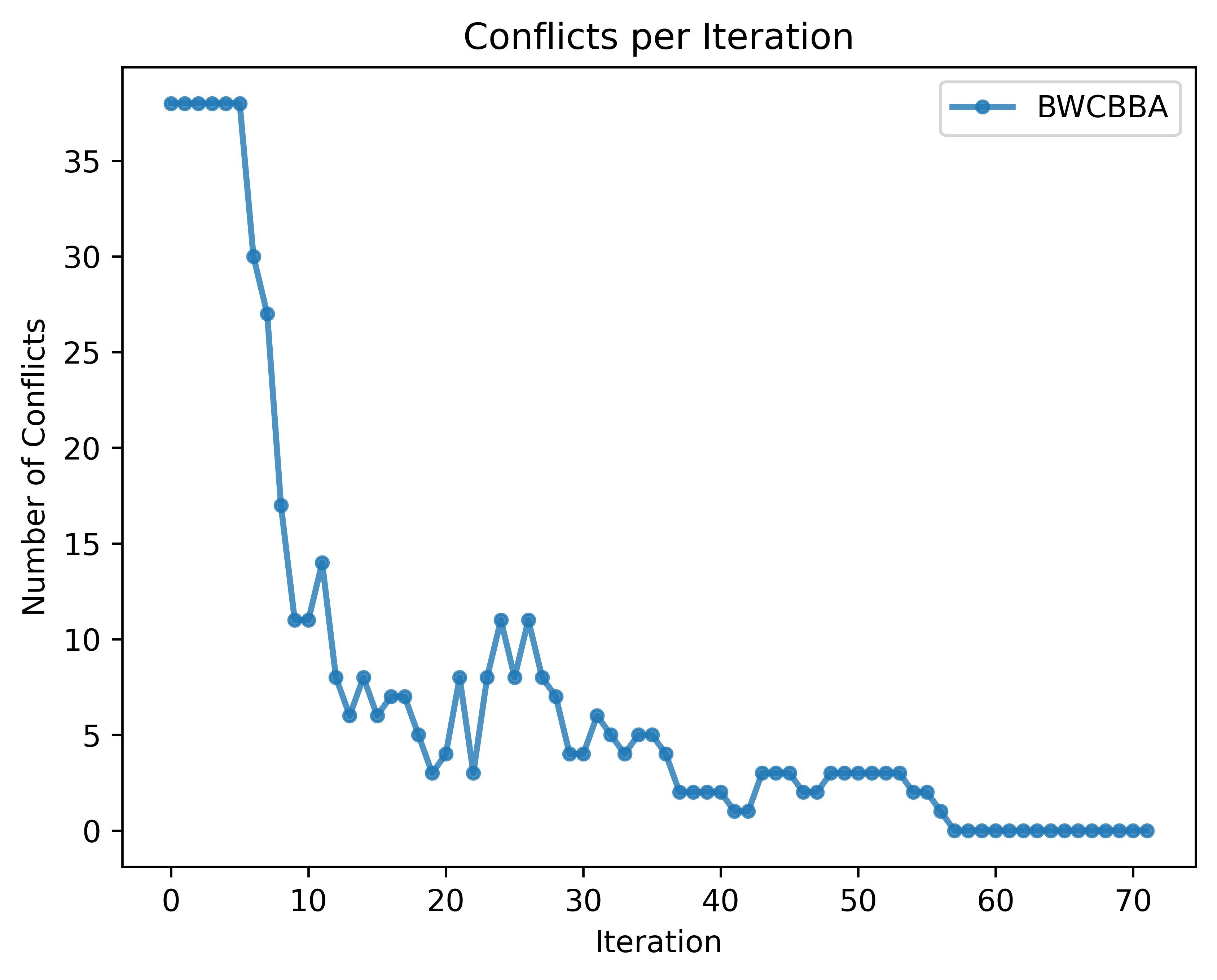}
        \caption{Conflicts per BWCBBA iteration.}
        \label{fig:conf}
    \end{minipage}%
    \begin{minipage}{.5\textwidth}
        \centering
        \includegraphics[width=1\linewidth]{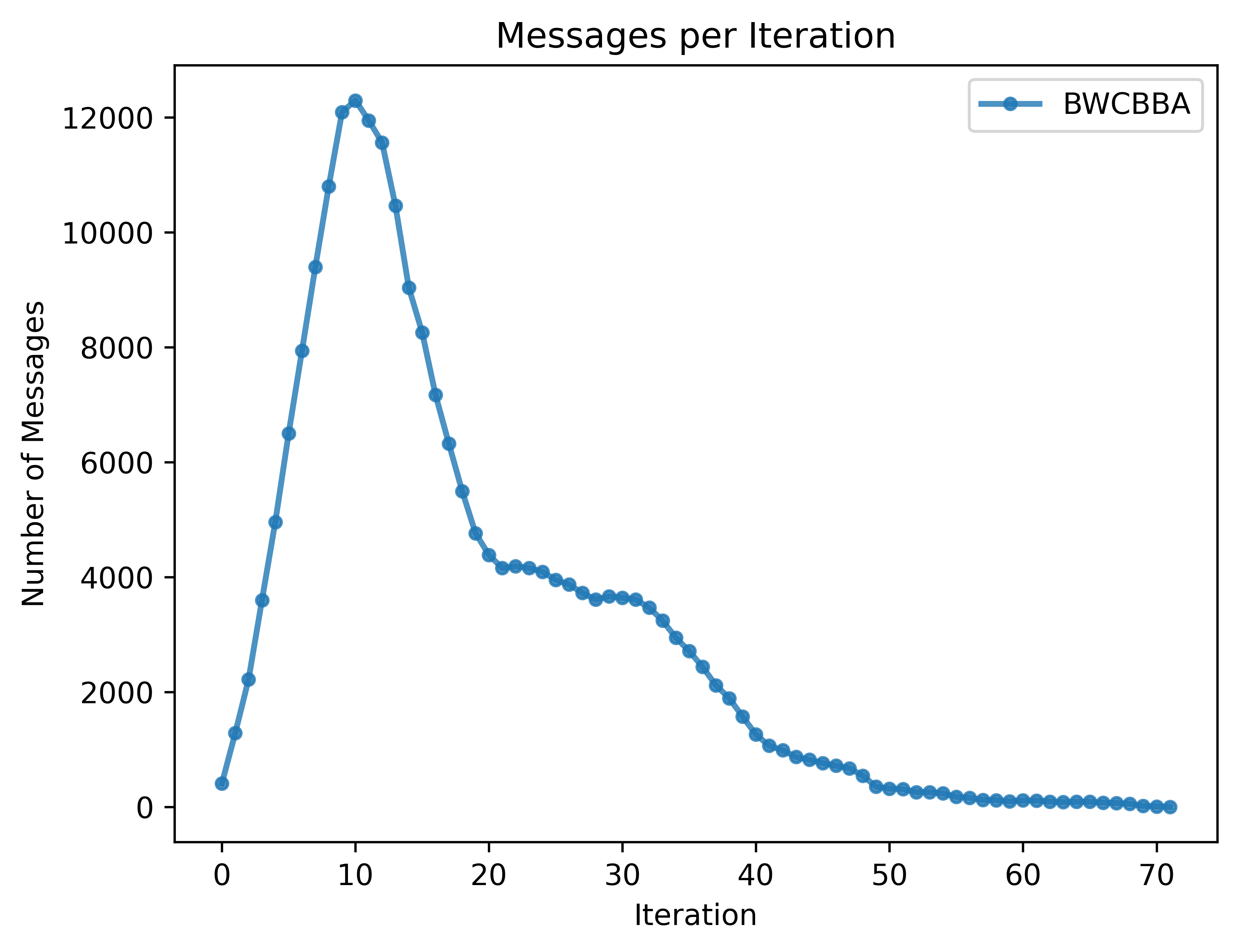}
        \caption{Messages per BWCBBA iteration.}
        \label{fig:msg}
    \end{minipage}
\end{figure}

Figure \ref{fig:conf} shows conflicts diminish rapidly, settling at zero around 57 iterations. Conflicts are counted when any two or more agents consider their own bid on a task the highest, resulting in non-unique task ownership. Figure \ref{fig:msg} depicts message exchange per iteration, with a peak at roughly 15 seconds corresponding to the steepest conflict decline. All spacecraft rebroadcast messages to neighbors, but only command agents generate bid messages. Communication is modeled as lossless between line-of-sight neighbors.

\begin{figure}
    \centering
    \begin{minipage}{.5\textwidth}
        \centering
        \includegraphics[width=1\linewidth]{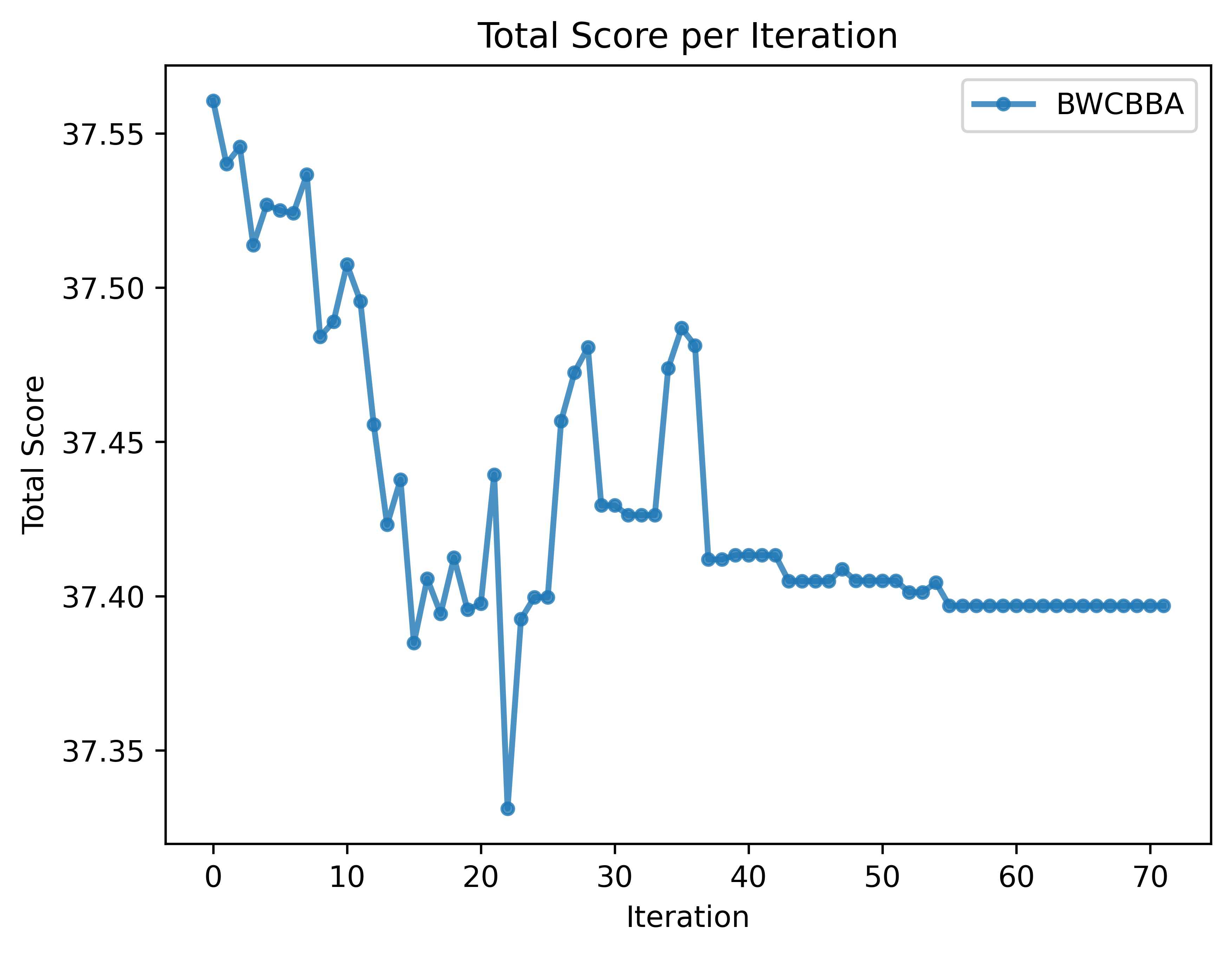}
        \caption{Global score per BWCBBA iteration.}
        \label{fig:score}
    \end{minipage}%
    \begin{minipage}{.5\textwidth}
        \centering
        \includegraphics[width=1\linewidth]{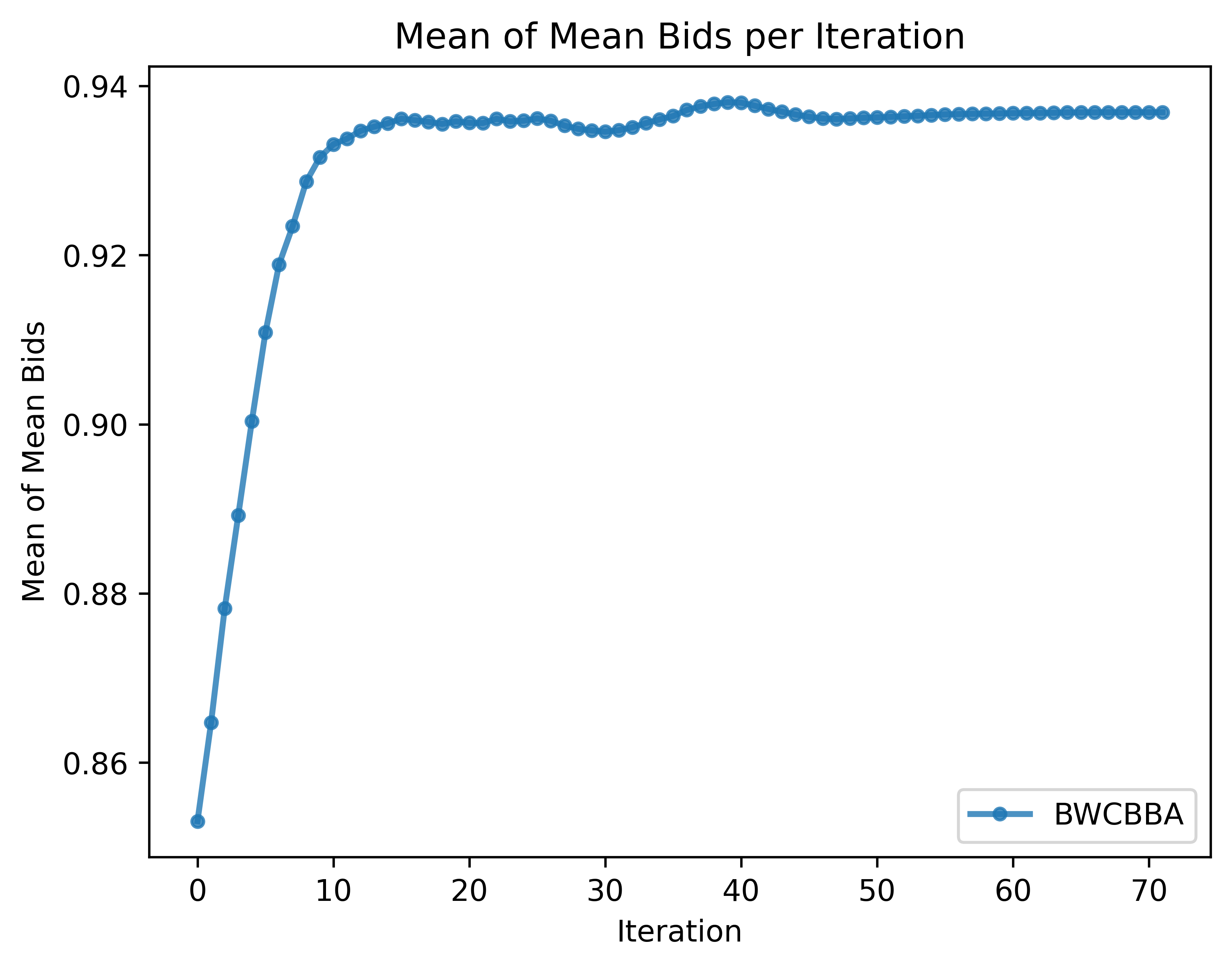}
        \caption{Mean bids per BWCBBA iteration.}
        \label{fig:bid}
    \end{minipage}
\end{figure}

Figure \ref{fig:score} shows total constellation performance across BWCBBA
iterations. Constellation performance is calculated using the diminishing returns function in Equation \ref{eq:reward}, and observation values calculated using the specific pointing assigned by the scheduler. Performance varies but remains within 1\% of the centralized score (37.5312), demonstrating that single target ownership and decentralization imposes negligible performance loss despite the challenging scenario characteristics.

Figure \ref{fig:bid} shows the mean of task bid means per iteration, calculated by averaging all bids for each task across agents' local bid spaces. Bid values increase per iteration as the auction allocates targets to higher-value agents, settling quickly with strong negative correlation to conflicts (Figure \ref{fig:conf}). The positive trend in bid values follows expected auction algorithm behavior despite the stagnant performance seen in the score evolution per iteration.

\subsection{Fusion Performance}
To analyze fusion performance, we first look at the completeness of the target tracking. As defined in Single Integrated Air Picture (SIAP) metrics\cite{votrubaSingleIntegratedAir}, completeness is the fraction of true targets being tracked. Table \ref{tab:fusion_completeness} aggregates completeness over the course of the simulation where a target is considered tracked if a track exists within 5 kilometers of the true target position at the sample time. As expected, the expanded FOV of the fusion-only scenario simplifies the problem geometry and thus provides results beyond what the tasking scenarios can achieve. For the limited FOV scenarios, the second scenario provides a performance baseline that the decentralized fusion architecture in the third scenario nearly matches. The full decentralization of the system has degraded performance and is investigated in more detail with other metrics.

\begin{table}[]
    \centering
    \begin{tabular}{c|c|c|c}
        Test \# & Tasking       & Fusion        & Completeness (5 km) \\\hline
        1       & Nadir         & Decentralized & 0.873               \\
        2       & Centralized   & Centralized   & 0.364               \\
        3       & Centralized   & Decentralized & 0.347               \\
        4       & Decentralized & Decentralized & 0.192
    \end{tabular}
    \caption{Aggregate target completeness}
    \label{tab:fusion_completeness}
\end{table}

\begin{figure}[h!]
    \centering
    \includegraphics[width=1.0\textwidth] {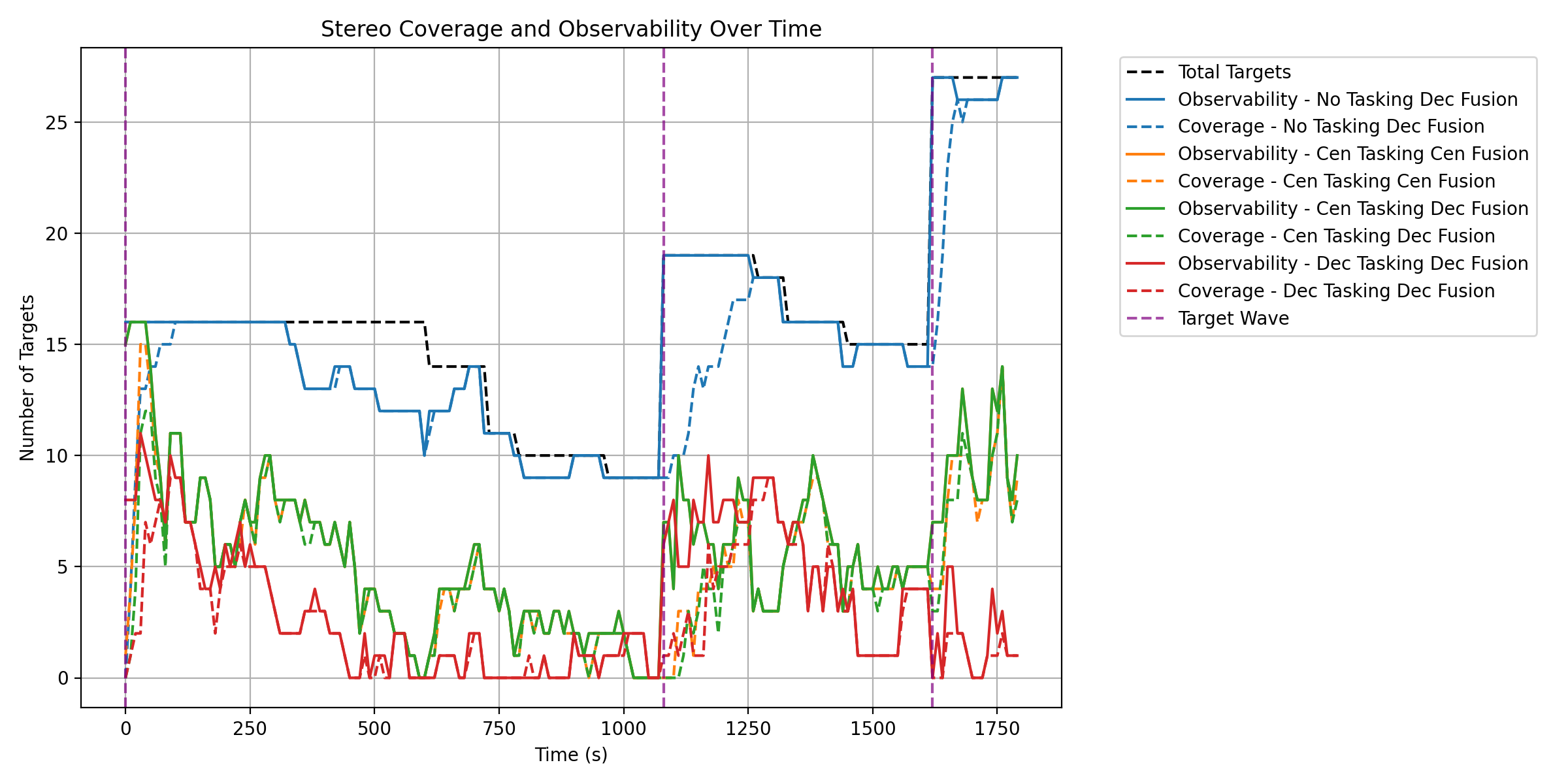}

    \caption{Stereo coverage and observability over time.}
    \label{fig:coverage}
\end{figure}

We next examine the stereo observability and coverage of the targets throughout the scenario. We define stereo observability to be the number of targets concurrently observed by at least two sensing satellites and stereo coverage to be the number of targets that have a track which has recently incorporated at least two sensing satellites' measurements of that target. It follows that the upper bound of coverage will be the observability. Figure \ref{fig:coverage} shows a general trend of the fusion system frequently matching stereo observations with stereo coverage. In the fusion-only case, the scenario generally has full stereo observability of the targets. The stereo coverage deviates from the observability primarily in the target wave times as new tracks are initialized; this is most noticeable after the second target wave. This system response can be seen in each scenario with slightly shorter response times from the centralized fusion scenario; while the decentralized response times are comparable, the full set of measurements ultimately brings about stereo initializations quicker. The fully decentralized scenario's performance is better examined here; while the coverage was generally able to meet the observability, the low trends in observability ultimately impacted the overall completeness seen in Table \ref{tab:fusion_completeness}.

\begin{figure}[h!]
    \centering
    \includegraphics[width=1.0\textwidth] {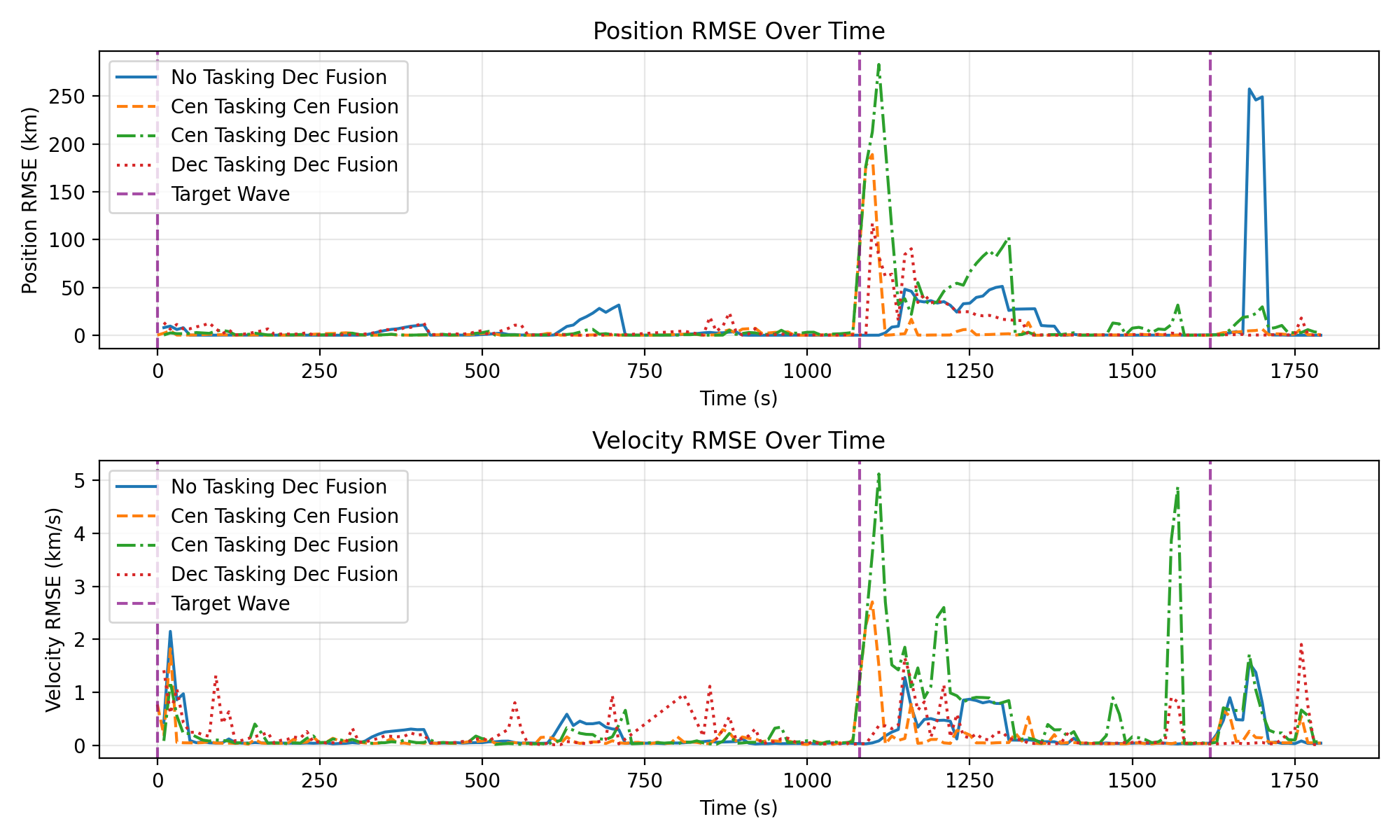}

    \caption{RMSE over time.}
    \label{fig:rmse_metrics}
\end{figure}

We use the root mean squared error (RMSE) to examine the positional and velocity estimates of the tracks. Track-to-truth assignments are performed by simply minimizing each track's error against possible associations. In Figure \ref{fig:rmse_metrics}, the scale of the plot is dominated by association errors; each scenario experiences either poor initializations or tracks that drift from targets which results in spikes in the RMSE plot. Regardless of observability, each scenario is able to achieve consistent low RMSE values throughout the simulation. The worst values occur after the second target wave introduction, matching the slower system response for stereo coverage seen in Figure \ref{fig:coverage}.

\begin{figure}[h!]
    \centering
    \includegraphics[width=1.0\textwidth] {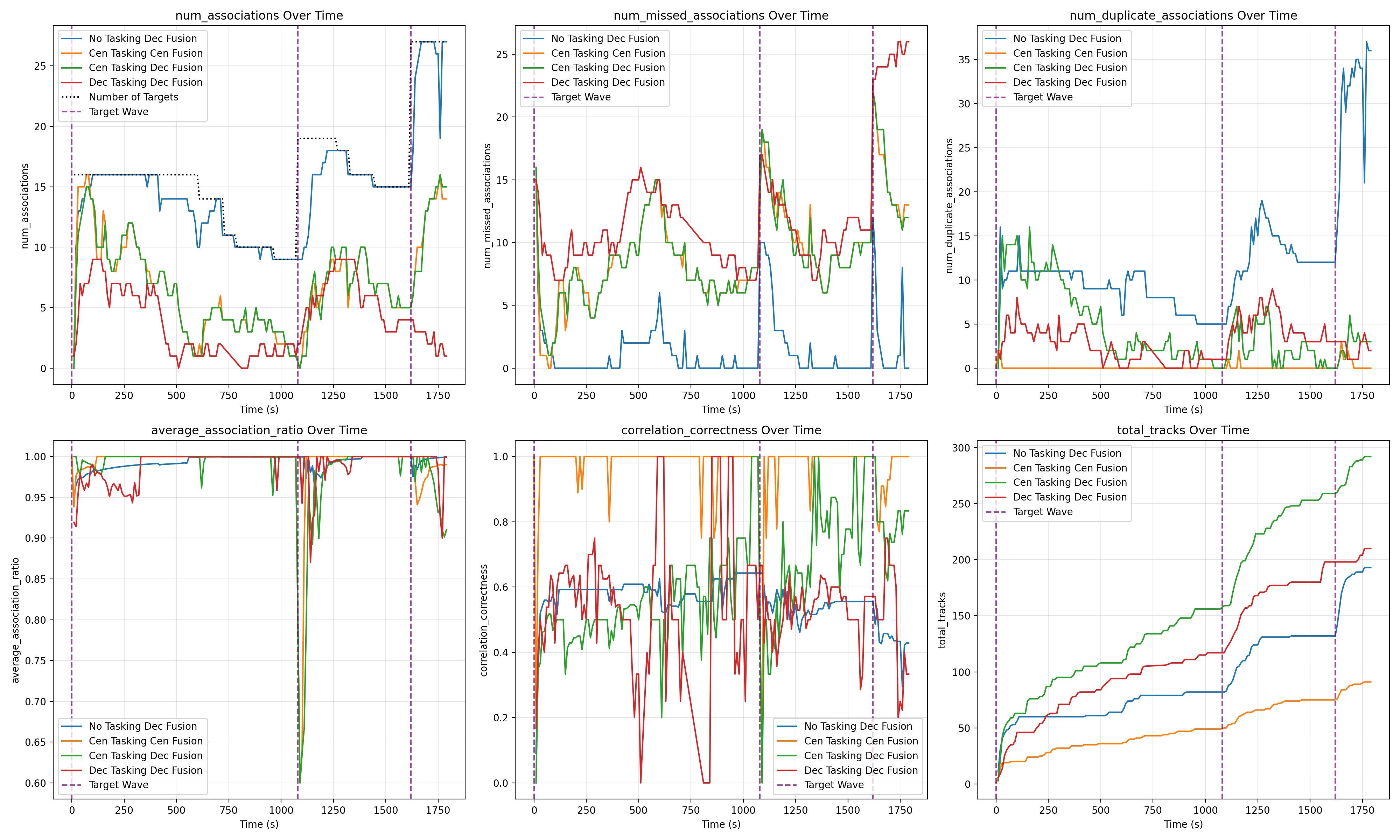}

    \caption{Association metrics.}
    \label{fig:measurement_metrics}
\end{figure}

In Figure \ref{fig:measurement_metrics}, we further characterize each architecture's performance with an array of metrics related to the association of tracks to targets. For these metrics, we match track-to-truth associations by examining the ground truth target that corresponds to the measurement update of each estimate. We first observe that the number of associations and missed associations closely matches the stereo coverage results. The centralized fusion architecture, reasoning over the full set of measurements, creates a relatively low number of duplicate tracks and quickly resolves them. The decentralized fusion architecture, introducing duplicate information through the geocasting and working with partial information, creates more duplicates and resolves them slower throughout the course of the simulation.

In the bottom row of Figure \ref{fig:measurement_metrics}, we consider the association ratio as the ratio of a track's measurements from the target it is predominately tracking to the total number of measurements incorporated into the track. This measure stays consistently high across all methods but drops considerably in the response to the second target wave. Following Ogle et al\cite{ogleAssessmentHierarchicalMultisensor2019}, we measure correlation correctness as the following:

\begin{equation}
    CC_k = \frac{N_s - N_m - N_d}{N_s}
\end{equation}

In this notation, $N_s$ is the number of system tracks, $N_m$ is the number of misassociated tracks with recent information from multiple targets, and $N_d$ is the number of duplicate tracks. As expected, the centralized fusion scenario has much higher correlation correctness as it produces fewer duplicate tracks. The fusion-only scenario keeps a relatively stable result around $0.5$ while the other scenarios vary considerably as the observability changes.

Lastly, we plot the total number of tracks created throughout the simulation. The centralized fusion architecture creates the smallest number of tracks as it creates fewer duplicates. Despite the lower coverage, the decentralized fusion scenarios with tasking generate more tracks than the fusion-only scenario, suggesting that these scenarios generally struggled to keep a stable, long-duration track on targets.

\section{Conclusion}
We present a simulation architecture for the investigation of distributed tasking and sensor fusion in a large constellation. Through simulation, we show that an end-to-end decentralized system to allocate and track targets shows degradation compared to a full FOV system, but relatively comparable performance to additional cases where sensing was limited by observation scheduling. Ongoing work seeks to reduce the deficits of the decentralized approaches, such as the number of duplicate tracks that are produced and the system's robustness to identifying and resolving them. In addition, efforts towards mitigating performance losses due to single target ownership in tasking will be investigated, with interest towards allowing shared target ownership and early exit criteria from consensus.

Other future work will look to evolve the system's design and algorithms to exploit the advantages of the decentralized approach. For instance, the system's capacity for low-latency responsiveness and the potential for resiliency to failures are future paths for research.

\bibliographystyle{AAS_publication}   %
\bibliography{references}   %

\end{document}